\documentclass{article} % For LaTeX2e
\usepackage{iclr2026_conference,times}
\usepackage[utf8]{inputenc}
\usepackage{multirow}
\usepackage[table]{xcolor}  % 加载 xcolor 并启用表格支持
\usepackage{colortbl}  
\usepackage{amsmath,amsfonts,bm}

\def\eqref#1{equation~\ref{#1}}
\def\1{\bm{1}}

\DeclareMathAlphabet{\mathsfit}{\encodingdefault}{\sfdefault}{m}{sl}
\SetMathAlphabet{\mathsfit}{bold}{\encodingdefault}{\sfdefault}{bx}{n}

\definecolor{iccvblue}{rgb}{0.21,0.49,0.74}

\usepackage[pagebackref,breaklinks,colorlinks,allcolors=iccvblue]{hyperref}
\usepackage{url}

\usepackage{amsmath}
\usepackage{booktabs}
\usepackage{multirow}
\usepackage{pifont}
\usepackage{bm}
\usepackage{graphicx}
\usepackage{subcaption}
\usepackage{wrapfig}

\title{\ModelName{}: A Framework for Integrating SID Redistribution and Length Reduction}

\author
{Zesheng Wang\thanks{Equal contribution.} \ , Longfei Xu\footnotemark[1] \thanks{Corresponding author.}\ , Weidong Deng, \\
\textbf{Huimin Yan, Kaikui Liu, Xiangxiang Chu} \\
AMAP, Alibaba Group\\
}

\newcommand{\ModelName}{IntRR}
\iclrfinalcopy % Uncomment for camera-ready version, but NOT for submission.
\begin{document}

\maketitle

\begin{abstract}
Generative Recommendation (GR) has emerged as a transformative paradigm that reformulates the traditional cascade ranking system into a sequence-to-item generation task, facilitated by the use of discrete Semantic IDs (SIDs). 
However, current SIDs are suboptimal as the indexing objectives (Stage 1) are misaligned with the actual recommendation goals (Stage 2). Since these identifiers remain static (Stage 2), the backbone model \textbf{lacks the flexibility} to adapt them to the evolving complexities of user interactions.
Furthermore, the prevailing strategy of flattening hierarchical SIDs into token sequences leads to \textbf{sequence length inflation}, 
resulting in prohibitive computational overhead and inference latency. 

To address these challenges, we propose \ModelName{}, a novel framework that integrates objective-aligned \textbf{SID Redistribution} and structural \textbf{Length Reduction}.
By leveraging item-specific Unique IDs (UIDs) as collaborative anchors, this approach dynamically redistributes semantic weights across hierarchical codebook layers. Concurrently, \ModelName{} handles the SID hierarchy recursively, eliminating the need to flatten sequences. This ensures a fixed cost of one token per item. 
Extensive experiments on benchmark datasets demonstrate that \ModelName{} yields substantial improvements over representative generative baselines, achieving superior performance in both recommendation accuracy and efficiency.
\end{abstract}

\section{Introduction}

%传统序列推荐系统-->生成式推荐-->GR with SIDs
Traditional sequential recommendation models predominantly utilize item IDs as the sole representation of items, capturing collaborative information through historical user-item interactions~\citep{sasrec,din,dien,wang2017deep}. Leveraging the advanced natural language processing and multi-modal extraction capabilities, recent research has moved beyond simple ID-matching toward the GR paradigm~\citep{tiger,hstu,intsr,onerec}, reframing recommendation as a sequence-to-item generation task.

By quantizing multi-modal attributes into discrete token sequences, SIDs allow the recommendation task to be reframed as predicting the next item's structural tokens. 
This approach enables the model to capture complex user behaviors, while directly integrating content-aware relationships into the generative process.
Current GR models with SIDs follow a two-stage~\citep{onesurvey,sidsurvey,sidhandbook} decoupled paradigm: 
\textbf{(Stage 1) Semantic Indexing}, which constructs discrete item identifiers, and \textbf{(Stage 2) Generative Learning}, which trains the recommender to predict these identifiers. We argue that this disjoint optimization leads to representation and structural bottlenecks during the generative learning stage (Stage 2) (see Fig.~\ref{Figure: Schematic_overview}):

\begin{figure*}[t]
\centering
\includegraphics[width=0.85\textwidth]{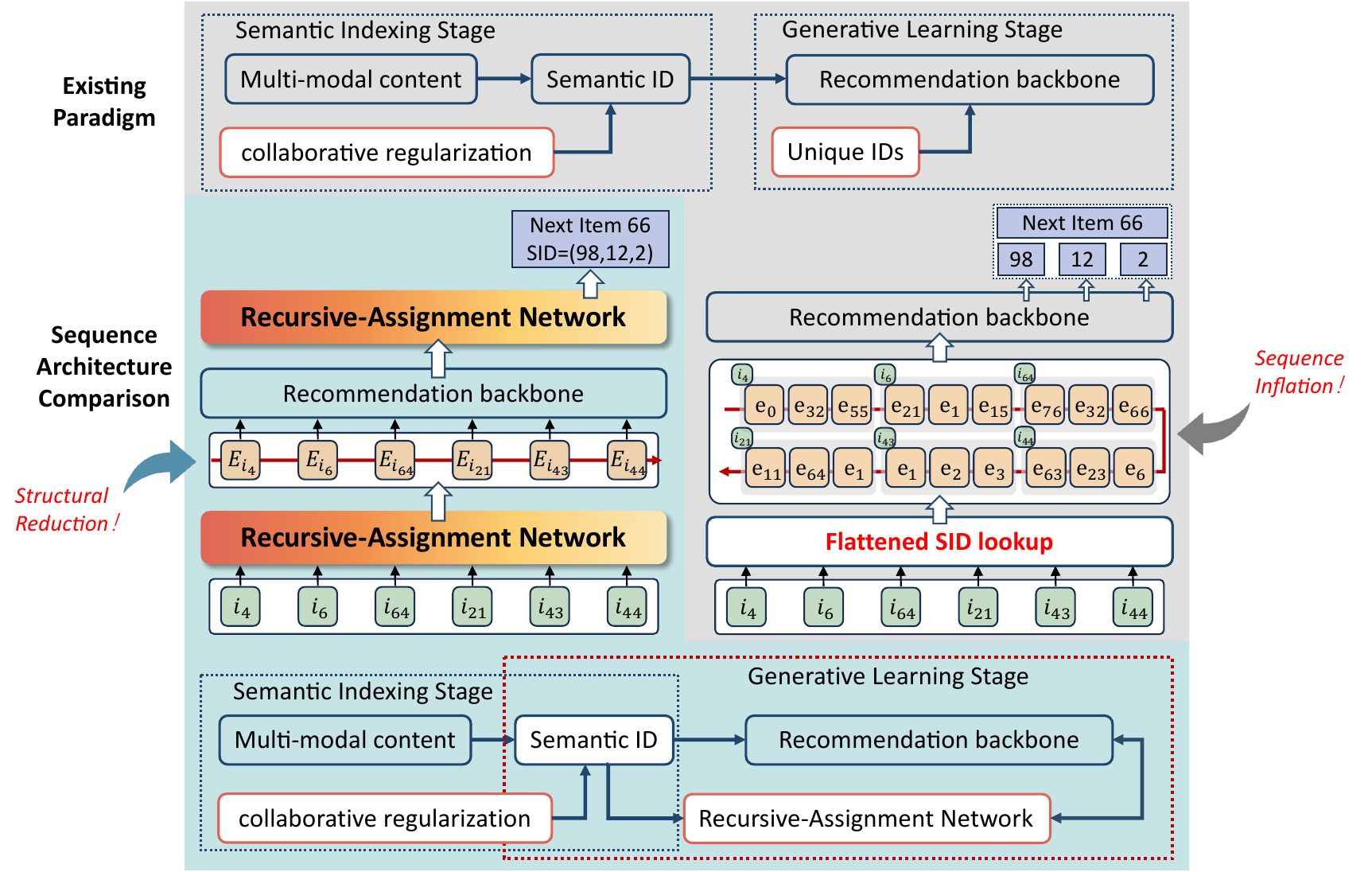}
\caption{Comparison of existing paradigms (gray) and \ModelName{} (blue). 
Row 1: Decoupled indexing (gray) yields suboptimal SIDs and representation ceilings, as identifiers remain task-agnostic despite incorporating collaborative regularization or UIDs. 
Row 2: Contrast between baseline sequence inflation ($L$ tokens per item, gray) and \ModelName{} reduction (1 token per item, blue)
Row 3: Our RAN achieves objective-aligned redistribution and deep collaborative-semantic integration at generative learning stage.}
\label{Figure: Schematic_overview}
\end{figure*}

Suboptimal identifier generation arises from the offline indexing stage (Stage 1).
Foundational models such as TIGER~\citep{tiger} and LIGER~\citep{liger} construct SIDs based solely on item content, completely overlooking collaborative interaction patterns. 
Although recent advances~\citep{letter,wang2024eager} attempt to bridge this gap by incorporating collaborative signals during indexing, these approaches cannot completely address the problem.

Since existing methods treat identifiers as static and rigid tokens (Stage 2), they encounter three significant challenges:
(i) \textbf{Constrained Item Representation}: 
Directly importing SIDs from the indexing stage restricts item embeddings to pre-allocated, frozen identifiers. 
When these identifiers conflict with nuanced user-item interaction patterns, the backbone lacks the flexibility to refine them, creating a representation ceiling.
(ii) \textbf{High Computational Complexity}: 
The prevailing strategy of flattening hierarchical SIDs into sequences significantly inflates the sequence length. (as shown in row 2 of Fig.~\ref{Figure: Schematic_overview})
Given the $O(N^2)$ complexity of Transformers, this leads to a ``computational explosion'' during training.
(iii) \textbf{Inference Inefficiency}: 
The step-by-step autoregressive nature of decoding SIDs incurs prohibitive latency, making real-time recommendation practically unfeasible in large-scale scenarios.

To address these challenges, we propose \textbf{\ModelName{}}, 
a novel framework that shifts the focus from optimizing offline indexing to enabling dynamic representation refinement and streamlined generation within the backbone.
At its core, \ModelName{} introduces a \textbf{Recursive-Assignment Network (RAN)}, which functions as a dynamic bridge during generative learning. 
RAN treats hierarchical SID navigation as a recursive task within the model backbone, thereby enabling objective-aligned redistribution and reducing structural length.  
More specifically, to break the representation ceiling (Issue i), 
RAN utilizes item-specific UIDs as collaborative anchors. These anchors adaptively redistribute importance weights across hierarchical codebooks. This mechanism allows the model to ``refine'' the coarse semantic priors from Stage 1. As a result, it generates collaborative-aware representations, which are directly optimized for the recommendation objective. 
Concurrently, the recursive design of RAN folds the hierarchical SID structure into internal hidden state transitions. 
%
% This restores the effective sequence length to a constant 1 token per item, successfully bypassing the quadratic complexity inflation during training (Issue ii) and the multi-step autoregressive overhead during inference (Issue iii).
%
This design mitigates quadratic training inflation (Issue ii) by maintaining a fixed one-token length for each item (see Fig.~\ref{Figure: Schematic_overview}). Simultaneously, it mitigates inference latency (Issue iii) by eliminating the step-by-step autoregressive bottleneck.

In summary, we summarize our contributions as follows:
\begin{itemize}
    \item \textbf{In-backbone Semantic Refinement:} We propose \ModelName{}, which enables dynamic SID redistribution via UIDs directly within the generative backbone.
    This mechanism reconciles the inherent conflict between static indexing and recommendation objectives, effectively breaking the representation ceiling of traditional SIDs.
    %
    %
    % \item \textbf{Efficiency-Centric Recursive Architecture:} We reformulate the SID-to-embedding mapping as an internalized \textbf{recursive task}. 
    % %
    % This design fundamentally eliminates the flattened-sequence paradigm and restores the per-item sequence cost to one token, 
    % %
    % successfully bypassing quadratic training complexity and replacing the multi-step SID autoregressive inference with a single-pass recursive prediction.
    %
    \item \textbf{Efficiency-Centric Recursive Architecture:} We reformulate the SID-to-embedding mapping as an internalized recursive task. This strategy reduces quadratic training costs via single-token representation and accelerates inference by skipping step-by-step generation.
    \item \textbf{Versatile Enhancement and Validation:} Extensive experiments on multiple benchmark datasets demonstrate that \ModelName{} serves as a robust enhancement module. %
    It yields consistent performance gains across various SID construction methods and backbone architectures, while substantially reducing the complexity of training and inference.
\end{itemize}  

\section{Related Work}
% ICLR的内容可以做补充
\subsection{Generative Recommendation}
GR has recently emerged as a transformative paradigm that reframes the traditional embedding-based retrieval as a sequence generation task~\citep{geng2022recommendation,lc-rec,climber,unisearch,lum}. As the first generative retrieval model, TIGER~\citep{tiger} employs quantized hierarchical SIDs and a sequence to sequence model to predict target item IDs based on user history. HSTU~\citep{hstu} replaces traditional numerical features with interleaved sequences of sparse item and action IDs. This unified design consolidates retrieval and ranking into a single generative task. IntSR~\citep{intsr} introduces a Query-Driven Block that decouples the processing of query placeholders from conventional historical behavior sequences, thereby achieving a unified framework for both search and recommendation. OneRec~\citep{onerec} first introduces an end-to-end generative architecture, effectively removing the need for a traditional multi-stage pipeline. IntTravel~\citep{inttravel} proposes the first multi-task generative recommendation framework, integrating four core tasks in the travel domain for end-to-end joint training.

\subsection{Item Representation via SIDs}

\subsubsection{Quantization and SID Construction}
The construction of SIDs typically involves a quantization process that maps continuous item content embeddings (derived from text or multi-modal encoders) into a hierarchical discrete space. Common techniques include Residual Quantization (RQ-VAE)~\citep{rqvae,tiger,lc-rec} Vector Quantised-Variational AutoEncoder (VQ-VAE)~\citep{tokenrec,universal} and Residual K-means (RK-Means)~\citep{onerec,zhai2025multimodal}. These methods allow items with similar content to share common prefix tokens, facilitating semantic-aware retrieval. However, a significant limitation of these quantization-based SIDs is that they are constructed via an offline reconstruction objective (i.e., minimizing the loss between original and quantized embeddings). This leads to a ``Static Bottleneck'': the identifiers are fixed before the recommendation model is trained. Since the quantization process is agnostic to the downstream recommendation task, the resulting SIDs may not be optimal for capturing user preferences, resulting in a representation that is suboptimal for actual recommendation performance.

\subsubsection{Integrating Collaborative Information}
To mitigate the content-only bias of SIDs, recent research has explored the integration of collaborative signals into the generative framework. Works such as EAGER~\citep{wang2024eager} and LETTER~\citep{letter} attempt to incorporate interaction patterns during the indexing stage by using pre-trained collaborative embeddings as supervision for the quantization process. While these methods introduce behavioral signals, the information alignment still cannot be guaranteed during the backbone training phase, due to their pre-obtained static IDs.

Furthermore, a pervasive challenge across existing SID-based models like TIGER~\citep{tiger} and LIGER~\citep{liger} is the Flattened-Sequence Paradigm. To model the hierarchy of SIDs, these methods flatten the layers of IDs into a long sequence of tokens. During training, this can lead to an increase in sequence length, resulting in a computational explosion. Furthermore, because it predicts step-by-step, inference time is also extended.

Unlike previous methods, our \ModelName{} solves these problems by moving from static indexing to dynamic redistribution. It uses a RAN to adjust the semantic codebook's weighting based on UIDs. Simultaneously, this avoids flattened sequences entirely by handling hierarchical assignments as a recursive internal process.

%
% 3.2 是“心脏”： 先把最创新的 RAN 模块讲清楚，它是后面两节的基石。
% 3.3 和 3.4 是“应用场景”： 分别对应物品怎么进（Encoding）和推荐怎么出（Decoding）。这完美对应了Pipeline 图中的左侧支路和右侧支路，且强调了参数共享。
% 闭环优化： 3.5 节将所有模块统一在两个 Loss 之下，回应了开头提到的 Objective Misalignment
%

\subsection{Decoding Acceleration Techniques}

GR Systems often face inference-stage latency due to the quadratic complexity of sequential token-wise decoding. Addressing this, various acceleration techniques for autoregressive decoding have emerged\citep{forge,mtgr,hou2025generating}. A key approach in this direction is the dynamic pruning of the search space according to partial decoding results. For instance, beam search\citep{beamsearch} retain only a small number of candidate sequences, thereby preemptively discarding unpromising computational paths during multi-step generation.

\section{Methodology}
%
% 总起段，给出所有3.x章节的叙述逻辑
In this section, we present the technical details of \ModelName{}. 
To address the representation and efficiency bottlenecks in existing GR paradigms, we first establish the formal problem formulation and preliminaries in Sec.~\ref{Sec: Formulation}. 
The RAN is a shared recursive module, and we describe it in detail in Sec.~\ref{Sec: RAN}.
We then detail the dual-role application of RAN in two synergistic scenarios: 
Sec.\ref{Sec: Redistribution} describes the adaptive semantic weight redistribution for refining item representations, 
while Sec.\ref{Sec: Prediction} illustrates how it achieves length-reduced generative prediction for efficient recommendation. 
Finally, Sec.\ref{Sec: Training} presents the joint optimization strategy designed to integrate collaborative behaviors and semantic structures.

\subsection{Problem Formulation and Preliminaries}
\label{Sec: Formulation}
\begin{figure*}[t]
    \centering
    \includegraphics[width=0.99\textwidth]{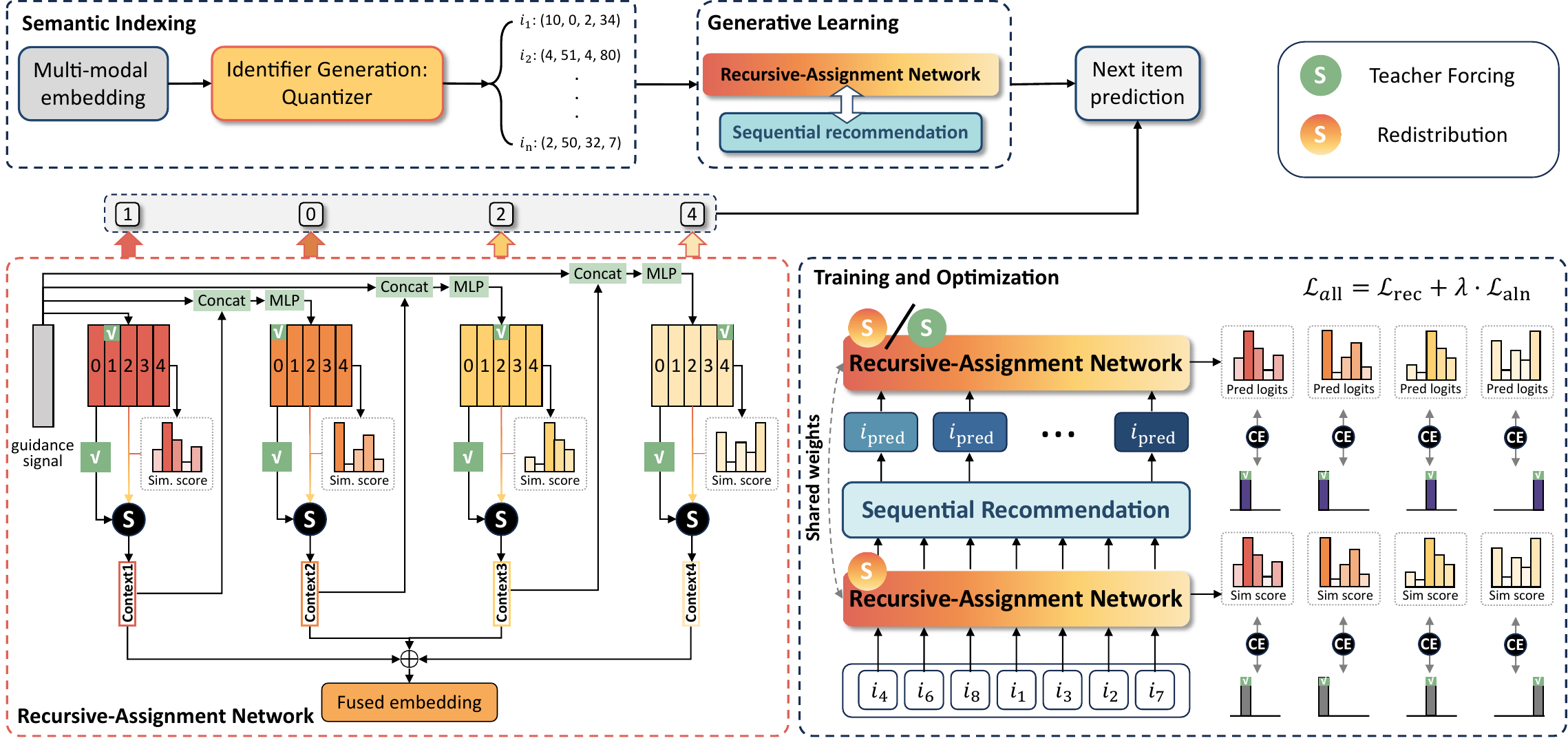}
    \caption{Overall architecture of \ModelName{}. The top pipeline illustrates the whole flow from semantic indexing to generative learning stage.
    The bottom-left panel details the RAN structure, featuring a dual-mode switch (Teacher Forcing vs. Redistribution). 
    While the bottom-right panel visualizes the semantic alignment and the recommendation learning process, highlighting the joint optimization of alignment loss ($\mathcal{L}_{aln}$) and recommendation loss ($\mathcal{L}_{rec}$) via the balancing weight $\lambda$.}
    \label{Figure: Overall_framework}
\end{figure*}

We follow the prevalent two-stage decoupled paradigm for GR.
In Stage 1 (Semantic Indexing), each item $i \in \mathcal{I}$ is assigned a static hierarchical identifier $\mathbf{c} = [c_1, c_2, \dots, c_L]$ based on its content attributes. 
In Stage 2 (Generative Learning), a backbone model $\Theta$ is typically trained to predict these identifiers.
However, static SIDs from Stage 1 lead to a representation bottleneck. 
\ModelName{} reformulates Stage 2 by introducing the shared RAN, denoted as $G$, which serves as a unified engine to characterize the global semantic space.

\textbf{Unified Recursive Engine.} 
\ModelName{} operationalizes the hierarchical navigation of SIDs as a conditional joint probability through a recursive chain rule. 
Unlike existing paradigms that rely on the sequence cache for each step, this recursive process is internalized within the shared RAN.
Given any guidance signal $\mathbf{x}$, which represents either a latent item content or a user preference state, 
the likelihood of observing a SID sequence $\mathbf{c}$ is decomposed as:
\begin{equation}
    P(c_1, \dots, c_L \mid \mathbf{x}; G) = \prod_{l=1}^L P(c_l \mid c_{<l}, \mathbf{x}; G).
    \label{Eq: chain_rule}
\end{equation}
At each level, the shared $G$ determines the $l$-th SID distribution by conditioning on the signal $\mathbf{x}$ and the semantic context $c_{<l}$ propagated from preceding levels.
%
%Within this unified recursive engine, 
Specifically, \ModelName{} utilizes $G$ to map SIDs by varying the guidance signal $\mathbf{x}$:

\textbf{Item-Side Alignment:} 
We introduce UID embeddings $\mathbf{e}_{uid}$ as collaborative guidance signals to refine item representations. 
Here, the shared $G$ maps $\mathbf{e}_{uid}$ into the semantic space to derive a refined item representation $\mathbf{E}_i \in \mathbb{R}^d$, while using the static SID $\mathbf{c}$ as a structural prior to guide the process. The model maximizes the alignment likelihood:
\begin{equation}
    P_{aln} = P(\mathbf{c} \mid \mathbf{e}_{uid}; G).
    \label{Eq: aln_prob}
\end{equation}

\textbf{User-Side Recommendation:} 
For a user $u \in \mathcal{U}$ with history $\mathcal{S}_u$, the backbone $\Theta$ extracts a behavioral hidden state $\mathbf{h}_u$ from the sequence of refined item representations $\{\mathbf{E}_j\}_{j \in \mathcal{S}_u}$.
The shared $G$ then use $\mathbf{h}_u$ as the signal to map the user state back into the semantic space for next-item prediction, maximizing the recommendation likelihood:
\begin{equation}
    P_{rec} = P(\mathbf{c} \mid \mathbf{h}_u; G, \Theta).
    \label{Eq: rec_prob}
\end{equation}

% \textbf{Joint Optimization Goal.} 
% The final objective is to maximize the combined log-likelihood of both tasks, ensuring the learned latent space is both semantically grounded via $P_{aln}$ and behaviorally expressive via $P_{rec}$:
% \begin{equation}
%     \mathcal{L} = \sum_{i \in \mathcal{I}} \log P_{aln}(\mathbf{c} \mid \mathbf{e}_{uid}; G) + \sum_{u \in \mathcal{U}} \log P_{rec}(\mathbf{c} \mid \mathbf{h}_u; G, \Theta).
%     \label{Eq: total_objective}
% \end{equation}

\subsection{Recursive-Assignment Network}
\label{Sec: RAN}
The RAN is the core module of \ModelName{}, 
designed to handle hierarchical SIDs in a differentiable and non-flattened manner.

\textbf{Hierarchical Semantic Codebook.} 
The RAN $G$ incorporates $L$  hierarchical codebooks $\mathbf{V} = \{ \mathcal{V}_{1}, \dots, \mathcal{V}_{L} \}$,
where each codebook $\mathcal{V}_{l} \in \mathbb{R}^{K \times d}$ contains $K$ trainable basis vectors of dimension $d$. 
This unified structure aligns item representations and user predictions within a common global semantic space.

\textbf{Recursive Computational Cycle.} 
To internalize the hierarchical navigation, 
the RAN module executes a recursive, layer-by-layer mapping across the $L$ semantic levels of the SID. 
Specifically, given a guidance signal $\mathbf{x} \in \mathbb{R}^d$, which represents either a UID or a user behavioral state, 
the internal computation follows a three-step cycle at each level $l \in \{1, \dots, L\}$ (as shown in Fig.~\ref{Figure: Overall_framework}, bottom-left):

\textit{1) Hidden State Fusion:} 
The network integrates the guidance signal $\mathbf{x}$ with the semantic context $\mathbf{z}_{<l}$ propagated from preceding levels to derive the $\mathbf{h}_l$ for the $l$-th level SID distribution prediction:
\begin{equation}
    \mathbf{h}_l = 
    \begin{cases} 
    \mathbf{x}, & l=1, \\
     \text{MLP} [\mathbf{x} \parallel \mathbf{z}_{<l}], & 1 < l \le L,
    \end{cases}
    \label{Eq: RAN_fusion}
\end{equation}
where $[\cdot \parallel \cdot]$ denotes concatenation. 
The layer-specific $\text{MLP}$ transforms the fused collaborative-semantic information back into the $d$-dimensional latent space to capture deep interactions while maintaining structural consistency.

\textit{2) Hierarchical Token Assignment:} 
The hidden state $\mathbf{h}_l$ acts as a query to probe the $l$-th codebook $\mathcal{V}_{l}$, measuring the similarity score between the guidance signal and the $K$ semantic prototypes. The resulting assignment distribution $\mathbf{p}_l \in \mathbb{R}^K$ is computed as:
\begin{equation}
    \mathbf{p}_l = \text{Softmax} \left( \mathbf{h}_l \mathcal{V}_{l}^\top \right),
    \label{Eq: RAN_prob}
\end{equation}
where each element $p_l^{(k)} \in \mathbf{p}_l$ signifies the soft assignment probability of the signal $\mathbf{x}$ to the $k$-th semantic cluster at level $l$. 
% 说清楚P的含义
This distribution effectively provides a differentiable relaxation of the SID. The discrete token $c_l$ at the current level is pre-assigned SID.

\textit{3) Switchable Contextual Aggregation:} 
To propagate information to the subsequent level, the updated context $\mathbf{z}_l$ is determined via a mode-switching mechanism (denoted as ``S'' in Fig.~\ref{Figure: Overall_framework}).
This switch allows RAN to derive the semantic context in two distinct modes: 
(i) Redistribution mode, where codewords at $l$-th level are aggregated based on the assignment weights:
\begin{equation}
    \mathbf{z}_l^{soft} = \sum_{k=1}^K p_l^{(k)} \cdot \mathcal{V}_l^{(k)},
    \label{Eq: Context_Update}
\end{equation}
and (ii) Teacher-forcing mode, where the context is directly assigned as the basis vector corresponding to the target item's actual SID token at that level (i.e., $\mathbf{z}_l = \mathcal{V}_l^{(c_l)}$). 
Different operating modes can be selected depending on the modeling scenario.

The design of RAN (bottom-left of Fig.~\ref{Figure: Overall_framework}) enables \ModelName{} to facilitate (i) \textbf{Adaptive Redistribution} (Sec.~\ref{Sec: Redistribution}) by allowing collaborative signals to differentiably steer the semantic weights to derive refined item representations,
and (ii) \textbf{Length-Reduced Prediction} (Sec.~\ref{Sec: Prediction}) by offloading intra-item hierarchical modeling from the backbone to the internal recursive path. 

\subsection{Adaptive SID Redistribution}
\label{Sec: Redistribution}
% 现有方法的static lookup带来表征瓶颈
Given that SIDs are pre-obtained, the backbone training's reliance on static SID lookups that cannot adequately fit the recommendation objective. 
To overcome this limitation, \ModelName{} utilizes pre-assigned SIDs as general semantic anchors instead of fixed identifiers. Within these anchors, codebook weights at each level are dynamically redistributed, guided by collaborative signals through the RAN module. 
This mechanism ensures that the item’s representation is directly optimized through the recommendation task.

\textbf{Collaboratively Guided Redistribution.} 
For each item $i \in \mathcal{I}$, the UIDs embedding $\mathbf{e}_{uid} \in \mathbb{R}^d$ is utilized as the guidance signal in Fig.~\ref{Figure: Overall_framework} to drive the RAN module, allowing the redistribution process to be adaptively guided by latent interaction patterns. 
Specifically, at each level $l$, the hidden state $\mathbf{h}_l$ is derived by fusing the collaborative $\mathbf{e}_{uid}$ with the semantic context $\mathbf{z}_{<l}$ according to Eq.~(\ref{Eq: RAN_fusion}). 
This fused state then generates the differentiable assignment distribution $\mathbf{p}_l$ via Eq.~(\ref{Eq: RAN_prob}).
By computing the layer-wise representation $\mathbf{z}_l$ in soft-weighted mode (Sec.~\ref{Sec: RAN}) over $\mathcal{V}_{l}$ based on $\mathbf{p}_l$ via Eq.~(\ref{Eq: Context_Update}), the collaborative guidance facilitates a soft-refined semantic representation, transcending the rigidity of a simple hard lookup.

\textbf{Hierarchical Feature Fusion.} 
To capture the multi-scale semantic features, we consolidate these layer-wise contexts into a unified collaborative-semantic representation $\mathbf{E}_i$. Formally, $\mathbf{E}_i$ is synthesized via the integration of the hierarchical representations:
\begin{equation}
    \mathbf{E}_i = \sum_{l=1}^L \mathbf{z}_l.
    \label{Eq: Item_Synthesis}
\end{equation}
This adaptive representation $\mathbf{E}_i$ provides the recommendation backbone with a more expressive input, yielding a dynamically optimized representation tailored for the recommendation task.

\textbf{Semantic Grounding via Structural Anchors.} 
While the redistribution process provides essential expressivity, it must remain semantically coherent to avoid drifting into an unstructured latent space.
Accordingly, the pre-assigned SID $\mathbf{c} = [c_1, \dots, c_L]$ serves as a structural anchor. We maximize the log-likelihood of the ground-truth (GT) path under the collaboratively guided assignment distribution:
\begin{equation}
    \mathcal{L}_{aln} = - \sum_{i \in \mathcal{I}} \sum_{l=1}^L \log P(c_l \mid \mathbf{e}_{uid}, \mathbf{z}_{<l}),
    \label{Eq: Aln_Loss}
\end{equation}
where $P(c_l \mid \cdot)$ corresponds to the $c_l$-th probability of the assignment distribution $\mathbf{p}_l$ allocated to the GT token $c_l$.
By supervising these layer-wise soft weights $\mathbf{p}_l$ to center around the hard structural anchors, the alignment loss $\mathcal{L}_{aln}$ effectively constrains the collaborative-driven exploration within a predefined hierarchy.
This design ensures that the resulting representation is both behaviorally expressive and semantically consistent.

\subsection{Length-Reduced Generative Prediction}
\label{Sec: Prediction}

A major computational bottleneck in current GR frameworks is the sequence length inflation caused by SID flattening.
By expanding identifiers into $L$ discrete tokens, conventional models~\citep{plm,tiger,etegrec} force the Transformer backbone to process a sequence of $N \times L$ tokens for a user history of $N$ items. This expansion leads to quadratic complexity and prohibitive inference latency. 
\ModelName{} achieves Length-Reduced Generative Prediction by internalizing the intra-item hierarchical navigation within the recursive RAN module.

\textbf{Compact Sequence Modeling.} 
Instead of feeding flattened token sequences into the Transformer, we represent each item in the user's history $\mathcal{S}_u$ using the adaptive collaborative-semantic embedding $\mathbf{E}_i$ synthesized in Eq.~(\ref{Eq: Item_Synthesis}). This allows the input sequence length to remain at a constant length of one token per item.
By maintaining a compact sequence $\mathcal{S} = \{ \mathbf{E}_{i_1}, \dots, \mathbf{E}_{i_N} \}$, the Transformer backbone is unburdened from modeling the internal structure of SIDs, which facilitates the learning of extended behavioral interactions between items. 
The user's latent preference is then captured by the final hidden state of the backbone:
\begin{equation}
    \mathbf{h}_u = \text{Backbone}(\mathcal{S}).
\end{equation}

\textbf{Recursive Decoding and Prediction.} 
The user behavioral state $\mathbf{h}_u$ is passed to the shared RAN module, which now functions as a hierarchical decoder to map the latent preference back into the global semantic space. 
Specifically, we utilize $\mathbf{h}_u$ as the guidance signal in Fig.~\ref{Figure: Overall_framework} to drive the recursive process defined in Eq.~(\ref{Eq: RAN_fusion}). 
Notably, unlike the recursive mode used for item-side representation,
the RAN operates in the teacher-forcing mode (Sec.~\ref{Sec: RAN}) during training.
Specifically, at each recursive step $l$, the hidden state $\mathbf{h}_l$ is generated by fusing $\mathbf{h}_u$ with the GT context $\mathbf{z}_{<l}^{GT}$, where $\mathbf{z}_{<l}^{GT} = \mathcal{V}_{<l}^{(c_{l})}$ is obtained via a hard lookup of the previous codebooks on the target SID. Based on $\mathbf{h}_l$, the RAN module computes a probability distribution $\mathbf{p}_l$ over $\mathcal{V}_{l}$ to predict the $l$-th token of the target SID. The GR task is optimized via the cross-entropy loss:
\begin{equation}
    \mathcal{L}_{rec} = - \sum_{(u, i) \in \mathcal{D}} \sum_{l=1}^L \log P(c_l \mid \mathbf{h}_u, \mathbf{z}_{<l}^{GT}),
    \label{Eq: Rec_Loss}
\end{equation}
where $\mathcal{D}$ denotes the set of user-item interactions. $P(c_l \mid \cdot)$ denotes the $c_l$-th codeword of the assignment distribution $\mathbf{p}_l$, representing the probability assigned to the target token $c_l$ given the user preference and the GT context. Through teacher forcing, predictions are strictly supervised by user preferences and the target SID’s hierarchical structure.

\textbf{Efficiency via Architectural Decoupling.} 
This recursive architecture fundamentally redefines the inference efficiency of GR. 
Traditional models~\citep{tiger} require $L$ successive Transformer forward passes to generate a single item SID, as each predicted token must be iteratively appended to the input sequence to predict the next. \ModelName{} only requires a single forward pass of the Transformer backbone to obtain $\mathbf{h}_u$, after which the entire SID sequence $[c_1, \dots, c_L]$ is decoded within the lightweight RAN module. 
By offloading the hierarchical navigation from the backbone to the efficient recursive unit, we eliminate the prohibitive inference latency of flattened sequences while maintaining advanced performance.

\subsection{Training and Optimization}
\label{Sec: Training}

To unify collaborative behaviors with semantic structures, the joint optimization objective (bottom-right of Fig.~\ref{Figure: Overall_framework}) is formulated as:
\begin{equation}
    \mathcal{L}_{total} = \mathcal{L}_{rec} + \lambda \mathcal{L}_{aln},
    \label{Eq: Total_Loss}
\end{equation}
where the alignment loss $\mathcal{L}_{aln}$ defined in Eq.~(\ref{Eq: Aln_Loss}) acts as a structural constraint, while 
$\mathcal{L}_{rec}$ defined in Eq.~(\ref{Eq: Rec_Loss}) serves as the primary recommendation objective.
The balancing weight $\lambda$ is a hyperparameter that modulates the trade-off between the semantic alignment and the recommendation task. This joint optimization objective ensures that \ModelName{} simultaneously preserves item's semantic structure while capturing the user's collaborative dynamics.
%To ensure stable convergence, we implement a \textbf{warm-up and decay scheduler} for the balancing weight $\lambda$. During the initial warm-up phase, a higher $\lambda$ is assigned to prioritize the establishment of a robust semantic-collaborative mapping (alignment). In the subsequent fine-tuning phase, $\lambda$ is gradually decayed to allow the recommendation objective to dominate the optimization. This strategy ensures that the final model reconciles the structural rigidity of static indexing with the dynamic flexibility required for objective-aligned recommendation.

\section{Experiments}
\label{Sec: Experiments}
% 总起段，最后在re一遍，引用Grid
Following the experimental setup described in Sec.\ref{Sec: Experimental_Setup}, we evaluate \ModelName{} as a versatile plug-and-play framework. 
To verify its robustness, we integrate \ModelName{} into different backbones and evaluate its performance across various indexing strategies.
This allows us to verify how the proposed \textbf{SID Redistribution} and \textbf{Length Reduction} mechanisms simultaneously enhance recommendation accuracy and computational efficiency.
Specifically, we investigate the model's performance and generalization across diverse indexing methods and backbones (Sec.\ref{Sec: Overall_Performance}), analyze its training and inference efficiency (Sec.\ref{Sec: Efficiency}), and conduct in-depth ablation studies to validate the underlying mechanisms (Sec.\ref{Sec: Ablation}).

\subsection{Experimental Setup}
\label{Sec: Experimental_Setup}

\textbf{Datasets and Metrics.} 
We evaluate our model on three widely-used Amazon benchmarks: \textit{Beauty}, \textit{Sports}, and \textit{Toys} \citep{tiger, hstu}.
All datasets are processed following the standard 5-core filter protocol. 
We adopt the leave-one-out evaluation: the most recent interaction per user is used for testing, the second-to-last for validation, and the remaining sequence for training. 
To measure retrieval accuracy and ranking quality, we report Recall@K and NDCG@K for $K \in \{5, 10\}$ on the test set, using with the best validation Recall@10. Results are reported as the average of 5 independent runs.

\textbf{Implementation Details.} 
To ensure a fair comparison, we align all configurations between \ModelName{} and the baseline within the SID-based GR paradigm.
For Semantic Indexing Stage, identical quantization methods are applied to generate the initial SIDs, including RK-Means \citep{onerec}, VQ-VAE \citep{vqvae}, and RQ-VAE \citep{rqvae}. All strategies generate hierarchical SIDs with a depth of $L=4$ (representing 4 discrete code layers) and a codebook size of 
$K=128$ at each level.
Regarding the generative learning stage, \ModelName{} and the baseline utilize identical backbones: the T5-based Transformer and HSTU \citep{hstu}. Both architectures consist of 4 layers. 
Specifically, the T5-based Transformer uses a hidden dimension of 256 and 6 attention heads, while HSTU employs a hidden dimension of 128 and 4 attention heads.
The primary distinction is that \ModelName{} replaces the conventional SID flattening with the RAN as a plug-and-play framework.
For optimization, we employ the Adam optimizer with a learning rate of $10^{-3}$, a weight decay of $10^{-4}$, and a batch size of 512.
The alignment weight $\lambda$ is 0.1 tuned via grid search. 
All experiments are conducted on 8 NVIDIA L20 (48GB) GPUs.

\textbf{Evaluation Settings (SAC vs. SSL).} To demonstrate the efficiency of \ModelName{}, we conduct comparisons under two distinct settings:
\textbf{(1) Same Action Count (SAC):} Both paradigms are provided with the same maximum history length of $N$ items. This configuration rigorously evaluates how effectively each paradigm distills collaborative patterns from a fixed-length interaction history.
\textbf{(2) Same Sequence Length (SSL):} Both paradigms are restricted to the same maximum token budget in the backbone. By compressing each item into a single token, \ModelName{} theoretically perceives $L \times$ larger historical contexts than the flattening baseline, demonstrating superior efficiency in long-term modeling.

\begin{table*}[tb]
\centering
\caption{Overall performance comparison across various indexing strategies and backbone architectures. All models are evaluated under the \textbf{SAC setting} (historical sequences of up to 30 items). The best results for each indexing strategy are highlighted in \textbf{bold}. \textbf{Imp.} denotes the average relative improvement over the respective backbone across all four metrics.}
\label{Tab: Overall_Performance}
\small
\resizebox{0.99\linewidth}{!}{
\begin{tabular}{ll|cccc|c|cccc|c|cccc|c}
\toprule
\multirow{2}{*}{Indexing} & \multirow{2}{*}{Method} & \multicolumn{5}{c|}{Amazon Beauty} & \multicolumn{5}{c|}{Amazon Sports} & \multicolumn{5}{c}{Amazon Toys} \\ 
\cmidrule(lr){3-7} \cmidrule(lr){8-12} \cmidrule(lr){13-17}
& & R@5 & R@10 & N@5 & N@10 & \textbf{Imp.} & R@5 & R@10 & N@5 & N@10 & \textbf{Imp.} & R@5 & R@10 & N@5 & N@10 & \textbf{Imp.} \\ 
\midrule

\multirow{4}{*}{RK-Means} 
& Transformer & 0.0374 & 0.0553 & 0.0254 & 0.0311 & - & 0.0164 & 0.0256 & 0.0105 & 0.0135 & - & 0.0252 & 0.0426 & 0.0161 & 0.0217 & - \\
& + \ModelName{} & \textbf{0.0462} & \textbf{0.0640} & \textbf{0.0326} & \textbf{0.0384} & \textbf{+22.8\%} & \textbf{0.0185} & \textbf{0.0291} & \textbf{0.0129} & \textbf{0.0163} & \textbf{+17.5\%} & \textbf{0.0317} & \textbf{0.0479} & \textbf{0.0216} & \textbf{0.0268} & \textbf{+24.0\%} \\
\cmidrule(lr){2-17}
& HSTU & 0.0302 & 0.0469 & 0.0195 & 0.0248 & - & 0.0162 & 0.0225 & 0.0106 & 0.0127 & - & 0.0205 & 0.0332 & 0.0130 & 0.0171 & - \\
& + \ModelName{} & \textbf{0.0423} & \textbf{0.0612} & \textbf{0.0292} & \textbf{0.0353} & \textbf{+40.7\%} & \textbf{0.0162} & \textbf{0.0252} & \textbf{0.0111} & \textbf{0.0139} & \textbf{+6.5\%} & \textbf{0.0267} & \textbf{0.0406} & \textbf{0.0182} & \textbf{0.0227} & \textbf{+31.3\%} \\

\midrule

\multirow{4}{*}{VQ-VAE} 
& Transformer & 0.0393 & \textbf{0.0582} & 0.0269 & 0.0330 & - & 0.0149 & \textbf{0.0246} & 0.0099 & 0.0127 & - & 0.0219 & 0.0367 & 0.0140 & 0.0187 & - \\
& + \ModelName{} & \textbf{0.0408} & 0.0580 & \textbf{0.0292} & \textbf{0.0348} & \textbf{+4.4\%} & \textbf{0.0160} & 0.0244 & \textbf{0.0101} & \textbf{0.0129} & \textbf{+2.6\%} & \textbf{0.0284} & \textbf{0.0416} & \textbf{0.0195} & \textbf{0.0237} & \textbf{+27.3\%} \\
\cmidrule(lr){2-17}
& HSTU & 0.0318 & 0.0499 & 0.0215 & 0.0273 & - & \textbf{0.0151} & \textbf{0.0234} & \textbf{0.0096} & \textbf{0.0123} & - & 0.0184 & 0.0305 & 0.0115 & 0.0155 & - \\
& + \ModelName{} & \textbf{0.0367} & \textbf{0.0530} & \textbf{0.0265} & \textbf{0.0317} & \textbf{+15.3\%} & 0.0135 & 0.0210  & 0.0092 & 0.0116 & \textbf{-7.7\%} & \textbf{0.0217} & \textbf{0.0341} & \textbf{0.0150} & \textbf{0.0190} & \textbf{+20.7\%} \\

\midrule

\multirow{4}{*}{RQ-VAE} 
& Transformer & 0.0356 & 0.0510 & 0.0244 & 0.0294 & - & 0.0150 & 0.0227 & 0.0098 & 0.0122 & - & 0.0209 & 0.0326 & 0.0133 & 0.0170 & - \\
& + \ModelName{} & \textbf{0.0439} & \textbf{0.0623} & \textbf{0.0315} & \textbf{0.0375} & \textbf{+25.6\%} & \textbf{0.0153} & \textbf{0.0232} & \textbf{0.0101} & \textbf{0.0127} & \textbf{+2.9\%} & \textbf{0.0300} & \textbf{0.0447} & \textbf{0.0209} & \textbf{0.0256} & \textbf{+47.1\%} \\
\cmidrule(lr){2-17}
& HSTU & 0.0302 & 0.0453 & 0.0210 & 0.0259 & - & 0.0130 & 0.0196 & 0.0079 & 0.0100 & - & 0.0156 & 0.0239 & 0.0101 & 0.0128 & - \\
& + \ModelName{} & \textbf{0.0376} & \textbf{0.0536} & \textbf{0.0268} & \textbf{0.0320} & \textbf{+23.5\%} & \textbf{0.0152} & \textbf{0.0227} & \textbf{0.0110} & \textbf{0.0124} & \textbf{+24.0\%} & \textbf{0.0252} & \textbf{0.0372} & \textbf{0.0172} & \textbf{0.0211} & \textbf{+63.1\%} \\

\bottomrule
\end{tabular}
}
\end{table*}

\subsection{Overall Performance and Generalization}
\label{Sec: Overall_Performance}

Tab.\ref{Tab: Overall_Performance} summarizes the recommendation performance of \ModelName{} across various indexing strategies and backbone architectures. 
To ensure a rigorous comparison, all models are evaluated under the SAC setting, where both paradigms process identical historical sequences comprising up to 30 items.
Notably, while the baseline requires a 120-token sequence (i.e., $L \times N$) to represent these items, \ModelName{} utilizes a highly compressed budget of only 30 tokens.

%\subsubsection{Generalization across Indexing and Backbones}
As shown in Tab.\ref{Tab: Overall_Performance}, \ModelName{} demonstrates superior recommendation performance and remarkable robustness. By integrating \ModelName{} into the generative pipeline (\textbf{+ \ModelName{}}), significant gains are consistently achieved across all three semantic indexing schemes (RK-Means, VQ-VAE, and RQ-VAE) and both two backbones (Transformer and HSTU). Specifically, on the \textit{Amazon Toys} dataset with RQ-VAE indexing, \ModelName{} yields its most substantial average improvements: \textbf{63.1\%} for the HSTU backbone and \textbf{47.1\%} for the Transformer backbone. 
The robust compatibility with diverse indexing methods and backbone choices proves that 
the efficacy of the proposed RAN and the joint optimization strategy,
%\ModelName{} effectively refines the fundamental item representation and sequence structure, 
making it a versatile framework that can be seamlessly integrated into various existing generative recommendation pipelines.

\subsection{Efficiency Analysis}
\label{Sec: Efficiency}

In this section, we evaluate the computational advantages of \ModelName{} across both training and inference stages. 
By significantly reducing the effective sequence length and bypassing the autoregressive bottleneck, \ModelName{} addresses the primary scalability challenges of the SID-based GR paradigm.
Our analysis demonstrates that \ModelName{} achieves exceptional efficiency and scalability, delivering substantial gains in system throughput and memory efficiency while maintaining high recommendation effectiveness. 

\begin{table}[!h]
\centering
\caption{Training efficiency comparison under the \textbf{SAC setting} ($N=30, L=4$). Throughput is measured as Samples Per Second (SPS) on a single NVIDIA L20 GPU. \ModelName{} significantly outperforms the baseline by reducing the backbone sequence length from 120 to 30 tokens.}
\label{Tab: Efficiency_Train}
\small
\resizebox{0.66\linewidth}{!}{ 
    \begin{tabular}{llcc}
    \toprule
    Backbone & Method & Throughput (SPS) $\uparrow$ & Peak VRAM $\downarrow$ \\
    \midrule
    \multirow{3.5}{*}{Transformer} & Baseline (Flattened) & 24,381.0 & 9.70 GB \\
    & \textbf{\ModelName{} (Ours)} & \textbf{42,666.7} & \textbf{3.08 GB} \\
    \rowcolor[gray]{0.95} & \textit{Improvement} & \textbf{+75.0\%} & \textbf{-68.2\%} \\
    \midrule
    \multirow{3.5}{*}{HSTU} & Baseline (Flattened) & 34,133.3 & 2.85 GB \\
    & \textbf{\ModelName{} (Ours)} & \textbf{46,545.5} & \textbf{2.65 GB} \\
    \rowcolor[gray]{0.95} & \textit{Improvement} & \textbf{+36.4\%} & \textbf{-7.0\%} \\
    \bottomrule
    \end{tabular}
} 
\end{table}

\begin{table*}[t]
\centering
\caption{Inference latency (ms per batch) comparison across different beam widths $W$ on Amazon Beauty dataset. \ModelName{} consistently achieves the lowest latency across all search spaces, demonstrating superior computational efficiency by bypassing the multi-pass autoregressive bottleneck and offloading this hierarchical retrieval to the lightweight RAN.}
\label{Tab: Efficiency_Inference}
\small
\setlength{\tabcolsep}{8pt} % 略微增加间距使纯延迟表更美观
\resizebox{0.9\linewidth}{!}{
\begin{tabular}{llcccccc}
\toprule
\multirow{2}{*}{Method} & \multirow{2}{*}{Setting} & \multicolumn{6}{c}{Inference Latency (ms) $\downarrow$} \\ 
\cmidrule(lr){3-8}
& & $W=10$ & $W=15$ & $W=20$ & $W=25$ & $W=30$ & $W=50$ \\ 
\midrule
Baseline & SAC ($N=32$) & 42.8 & 59.1 & 80.4 & 118.6 & 139.9 & 232.7 \\
Baseline & SSL ($N=8$)  & 34.0 & 46.5 & 59.1 & 67.4 & 78.3 & 123.3 \\
\midrule
\textbf{\ModelName{}} & \textbf{SAC/SSL} ($N=32$) & \textbf{33.7} & \textbf{39.1} & \textbf{45.1} & \textbf{50.6} & \textbf{56.6} & \textbf{79.4} \\
\midrule
\rowcolor[gray]{0.95} \multicolumn{2}{l}{\textit{Speedup Ratio (vs. SAC)}} & \textbf{1.27$\times$} & \textbf{1.51$\times$} & \textbf{1.78$\times$} & \textbf{2.34$\times$} & \textbf{2.47$\times$} & \textbf{2.93$\times$} \\
\bottomrule
\end{tabular}
}
\end{table*}

\subsubsection{Training Efficiency}
We evaluate training efficiency using Throughput (Samples Per Second, SPS) and Peak VRAM Consumption as primary metrics. 

\textbf{Experimental Setup.} To ensure a fair comparison under identical information constraints, all experiments are conducted under the \textbf{SAC setting} ($N=30, L=4$). 
In this configuration, the conventional flattening baseline expands each item into $L$ tokens, resulting in a backbone sequence length of $N\times L=120$. 
In contrast, \ModelName{} utilizes the RAN to condense each item into a single latent token, maintaining a compact sequence length of $30$. 

\textbf{Throughput Analysis.} As illustrated in Tab.\ref{Tab: Efficiency_Train}, \ModelName{} achieves a significant boost in throughput across different backbones. Specifically, it improves the SPS of the standard Transformer by \textbf{75.0\%} and the highly-optimized HSTU by \textbf{36.4\%}. 
The disparity in gains stems from the inherent complexity of the architectures: the standard Transformer, characterized by $O((NL)^2)$ self-attention, is more sensitive to sequence length reduction than the more efficient HSTU. 

%V2:仅分析数据，不考虑backbone之间的优化程度存在diff
\textbf{VRAM Utilization.} \ModelName{} demonstrates a profound impact on memory conservation, particularly for standard Transformer architectures. For the Transformer backbone, the peak VRAM drops dramatically from 9.70 GB to \textbf{3.08 GB}, representing a \textbf{68.2\%} reduction. This massive saving is theoretically grounded in the reduction of sequence length from $NL$ to $N$, which linearly decreases activation storage and quadratically reduces the self-attention complexity from $O((NL)^2)$ to $O(N^2)$. 
While the HSTU backbone intrinsically exhibits high memory efficiency, \ModelName{} still provides an \textbf{7.0\%} reduction in VRAM by shortening the underlying sequence length. 

% V1：分析HSTU对效率本身有优化导致数据优化程度没有Transformer明显
% \textbf{VRAM Utilization.} \ModelName{} demonstrates a profound impact on memory conservation, particularly for standard Transformer architectures. For the Transformer backbone, the peak VRAM drops dramatically from 9.70 GB to \textbf{3.08 GB}, representing a \textbf{68.2\%} reduction. This massive saving is theoretically grounded in the reduction of sequence length from $NL$ to $N$, which linearly decreases activation storage and quadratically reduces the self-attention complexity from $O((NL)^2)$ to $O(N^2)$. 
% % 
% In contrast, while the HSTU backbone is intrinsically designed for activation memory efficiency (reducing per-layer states from $33d$ to $14d$ \citep{hstu}), \ModelName{} still provides a further \textbf{7.0\%} reduction in VRAM by shortening the underlying sequence length. 
% %
% Interestingly, \ModelName{} acts as an \textbf{``efficiency equalizer''}---it narrows the performance gap, bringing the memory footprint of a standard Transformer (3.08 GB) remarkably close to that of the highly optimized HSTU (2.65 GB). This observation confirms that \textbf{sequence length inflation} is the primary memory bottleneck in generative recommendation. By alleviating this, \ModelName{} empowers researchers to utilize versatile, standard backbones while achieving a memory budget previously only attainable by highly specialized architectures.

\subsubsection{Inference Latency and Scalability}
We evaluate the inference latency (ms/batch) across varying beam widths $W \in \{10, \dots, 50\}$ to assess real-time responsiveness.
Traditional GR models suffer from an autoregressive bottleneck, requiring $L$ sequential backbone passes to generate a complete SID. 
In contrast, \ModelName{} invokes the heavy Transformer backbone only \textbf{once} to predict a SID, offloading the hierarchical retrieval to the lightweight RAN.
%.

% 这里应该描述一下为什么不是接近4倍的提升，因为我们数据集分布原因
\textbf{Latency and Speedup.} As demonstrated in Tab.\ref{Tab: Efficiency_Inference}, \ModelName{} achieves a significant and growing speedup ratio compared to the baseline of the SAC setting. 
% 重点分析：随着 W 变大，优势变大
Notably, the speedup ratio increases from \textbf{1.27$\times$ ($W=10$)} to a substantial \textbf{2.93$\times$ ($W=50$)}.
This widening gap indicates that while the baseline suffers from the cumulative delays of multi-step beam search across $L$ layers, \ModelName{} maintains high efficiency by executing beam search over a single-step intent space.
Even when compared to the SSL baseline (which only processes 8 items, $N \times L=32$), \ModelName{} (processing 32 items) remains significantly faster across all beam widths. This confirms that our framework can support a much larger modeling capacity (4$\times$ history) while still staying within stricter latency constraints.

% \textbf{The ``Double Win'' Phenomenon.} Crucially, \ModelName{} achieves a \textit{double win} in both efficiency and effectiveness.
% %
% As shown in the performance columns of Table \ref{Tab: Efficiency_Inference}, at the maximum beam width ($W=50$), \ModelName{} not only provides the lowest latency (\textbf{79.4 ms}) but also achieves the highest accuracy, with a relative gain of \textbf{+12.2\% in R@10} and \textbf{+18.8\% in N@10} over the SAC baseline. 
% % 
% This result indicates that the condensed, collaborative-aligned representations provided by \ModelName{} are not only faster to decode but also more discriminative for the recommendation task. This Pareto improvement makes \ModelName{} highly suitable for production environments where both high throughput and high precision are paramount.

\subsection{Ablation Study and In-depth Analysis}
\label{Sec: Ablation}

In this section, we conduct extensive ablation experiments on the \textit{Amazon Beauty} dataset to validate the design choices and analyze how key components affect recommendation performance.

\subsubsection{Component-wise Ablation}
To assess the contribution of each core module, we compare the full \ModelName{} against four critical variants. 
Specifically, \textbf{w/o User-side TF} switches the RAN from the GT mode to the soft-weighted mode during user-side training, replacing the GT context $\mathbf{z}_{<l}^{GT}$ with predicted distributions.
\textbf{w/o $L_{aln}$} removes the alignment loss $\mathcal{L}_{aln}$, forcing the model to be optimized solely by the recommendation objective without semantic anchoring. 
\textbf{w/o Redistribution} switches the item-side RAN from the soft-weighted to the GT mode, where item representations are obtained via a simple lookup of the SID path instead of redistributed weights.
\textbf{w/o Shared RAN} utilizes independent RAN modules for the user and item sides instead of sharing a unified global semantic space.

\begin{table}[ht]
\centering
\caption{Main ablation results on Amazon Beauty dataset. We evaluate the contribution of each component across multiple top-$k$ metrics. The $\Delta$ column indicates the relative performance drop in NDCG@10 compared to the full model.}
\label{Tab: Ablation_Main}
\small
\resizebox{0.85\columnwidth}{!}{
\begin{tabular}{lccccc}
\toprule
Variant & Recall@5 & Recall@10 & NDCG@5 & NDCG@10 & $\Delta$ (N@10) \\
\midrule
\textbf{\ModelName{} (Full)} & \textbf{0.0462} & \textbf{0.0640} & \textbf{0.0326} & \textbf{0.0384} & - \\
\quad w/o User-side TF      & 0.0303 & 0.0416 & 0.0221 & 0.0257 & -33.07\% \\
\quad w/o $L_{aln}$       & 0.0414 & 0.0587 & 0.0285 & 0.0341 & -11.20\% \\
\quad w/o Redistribution    & 0.0377 & 0.0542 & 0.0259 & 0.0312 & -18.75\% \\
\quad w/o Shared RAN      & 0.0376 & 0.0595 & 0.0266 & 0.0336 & -12.50\% \\
\bottomrule
\end{tabular}
}
\end{table}

\begin{figure}[h]
    \centering
    \includegraphics[width=0.92\columnwidth]{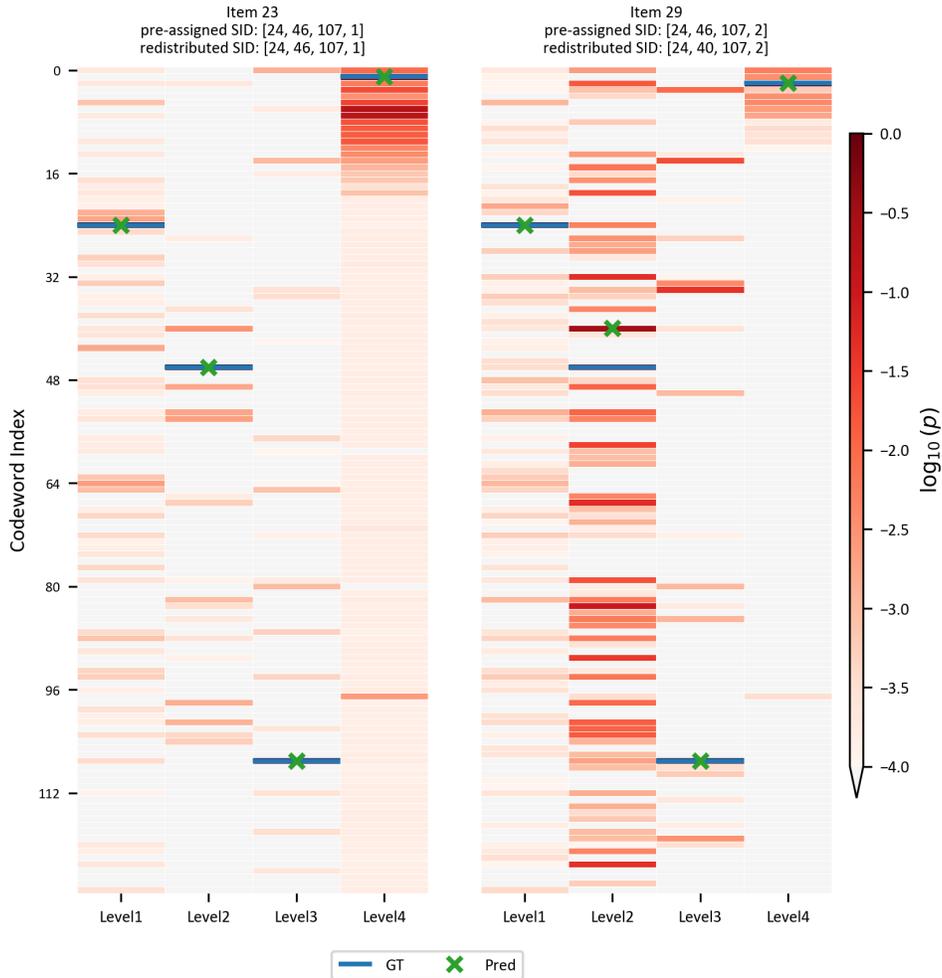} 
    \caption{Visualization of weight distribution for two items with the same SID.
    The RAN effectively achieves the redistribution of weights based on collaborative signals, yielding a more refined and unique representation for Item 23 and triggering a distinct semantic shift at level 2 for Item 29.}
    \label{Fig: Visualization}
\end{figure}

\textbf{Analysis.} As summarized in Tab.\ref{Tab: Ablation_Main}, \ModelName{} consistently outperforms all variants. 
\textbf{(1) Training Stability:} The \textbf{-33.07\%} drop in NDCG@10 without User-side TF confirms that stable GT signals are indispensable for mitigating error propagation during recursive decoding. 
\textbf{(2) Semantic Anchoring:} The \textbf{11.20\%} decay in \textit{w/o $L_{aln}$} validates hierarchical SIDs as vital structural priors, without which the model fails to distinguish complex item relationships. 
\textbf{(3) Value of Collaborative Refinement:} The \textbf{18.75\%} loss in \textit{w/o Redistribution} proves that simple lookups of static semantic centroids are suboptimal. Dynamic redistribution allows the RAN to inject collaborative nuances into content-based SIDs.
As visualized in Fig.~\ref{Fig: Visualization}, for items with identical SID, the redistribution exhibits a sophisticated transition:
for Item 23, while the peak probabilities remain aligned with the GT, the RAN assigns weight across other codewords at each level, yielding a more expressive and unique item representation.
For Item 29, the collaborative guidance triggers a more pronounced semantic weight shift at the second level, where the dominant weight migrates to a different codeword. This demonstrates how interaction-driven signals adaptively steer the representation away from static content-based priors to better align with recommendation goals. 
Since level 4 serves as a discriminator for items with identical prefixes (where unique SID default to index 0 and colliding ones are assigned sequentially), the weight distributions at this level are still predominantly concentrated within the initial codeword indices.
\textbf{(4) Value of a Unified Latent Space:} The \textbf{12.50\%} drop in \textit{w/o Shared RAN} confirms that a shared RAN aligns user and item representations within a unified domain, thereby enhancing matching effectiveness.

% \subsubsection{Impact of Semantic Anchoring Weight $\lambda$}
\subsubsection{Impact of Semantic Anchoring Weight \texorpdfstring{$\lambda$}{λ}}

To investigate the balance between structural anchoring and redistribution flexibility, we evaluate \ModelName{} across a range of $\lambda \in [0.0, 5.0]$. 

\begin{figure}[t]
    \centering
    \includegraphics[width=0.7\columnwidth]{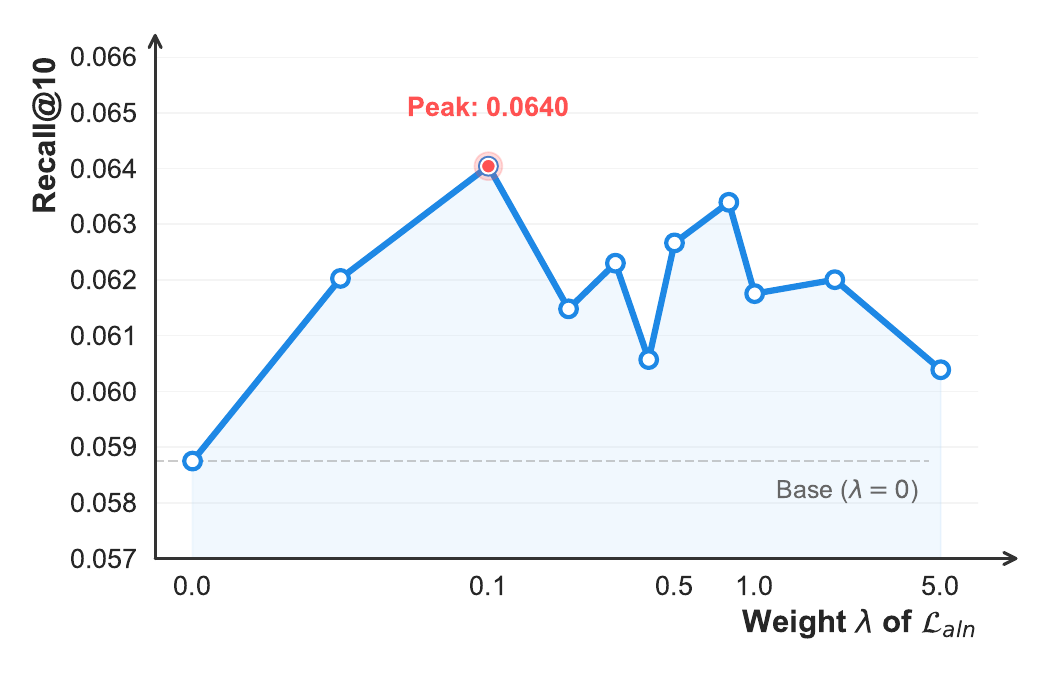}
    \caption{Sensitivity analysis of the weight $\lambda$ of $\mathcal{L}_{aln}$ on Beauty dataset. The performance peaks at $\lambda=0.1$, illustrating the optimal trade-off between hierarchical structural anchoring and collaborative-driven redistribution.}
    \label{Fig: Lambda_Sensitivity}
\end{figure}

\textbf{Analysis.} As illustrated in Fig.~\ref{Fig: Lambda_Sensitivity}, the performance exhibits a clear rise-then-fall trend as $\lambda$ increases, revealing how $\lambda$ modulates the ``degree of freedom'' in SID redistribution:
\textbf{(1) Excessive Freedom ($\lambda < 0.1$):} Performance starts at its lowest point ($0.0587$ at $\lambda=0$). When $\lambda$ is too small, the RAN possesses excessive freedom to relocate item representations with minimal structural constraint from the SID GT. This lack of guidance leads to severe ``semantic drift'', where the collaborative signals fail to ground within a stable hierarchy, resulting in suboptimal performance.
\textbf{(2) Optimal Balance ($\lambda = 0.1$):} Recommendation accuracy peaks at $\lambda=0.1$.
This value achieves an optimal balance between the fixed structure of SIDs and the flexibility of dynamic redistribution.
At this point, the structural guidance from $\mathcal{L}_{aln}$ effectively constrains the latent space while allowing the RAN enough freedom to adapt item representations to collaborative patterns.
\textbf{(3) Excessive Constraint ($\lambda > 0.1$):} 
As $\lambda$ increases further, performance fluctuates but follows a general downward trend, decreasing to 0.0603 at $\lambda = 5$. 
The growing structural constraint suppresses the redistribution freedom, making representations too rigid. This semantic rigidity limits the model's ability to adapt to complex user-item dynamics.

%
% \subsubsection{Probabilistic Soft-weighted TF for Autonomous Modeling}
% During inference, \ModelName{} must rely entirely on its own predicted representations without any GT guidance. However, a model trained exclusively with Teacher Forcing ($P=1.0$) on the user-side history tends to develop ``training inertia,'' where it over-relies on high-quality GT signals and fails to generalize when faced with its own imperfect predictions at test time. To mitigate this exposure bias, we introduce a \textbf{Probabilistic Soft-weighted TF} strategy. Specifically, we define $P$ as the probability of using the GT semantic centroid for historical actions, while with a probability of $1-P$, the model is forced to use its own \textit{softmax-weighted sum} of codebook basis vectors (generated by the RAN) as input.

% As illustrated in Table \ref{Tab: Ablation_P}, we analyze the performance across varying values of $P \in \{0.0, 0.2, 0.5, 0.8, 1.0\}$. The results reveal a clear trade-off: when $P$ is too small (e.g., $P=0.0$), the lack of stable supervision during the early stages of training hinders the convergence of the codebook. Conversely, when $P=1.0$, the model's performance degrades due to the discrepancy between training and inference (exposure bias). A moderate $P$ (e.g., $P=0.8$) proves to be the most effective, as it provides sufficient guidance to stabilize the latent space while simultaneously training the backbone to be robust against the ``soft'' and uncertain distributions it will encounter during real-world deployment.

\section{Conclusion}

In this paper, we proposed \ModelName{}, a framework that resolves the representation and efficiency bottlenecks in SID-based Generative Recommendation via the proposed RAN. 
By processing the SID hierarchy internally, the RAN enables dynamic \textbf{SID Redistribution} to refine coarse identifiers with collaborative nuances, while simultaneously achieving significant \textbf{Length Reduction} by eliminating sequence inflation. 
Extensive experiments demonstrate that \ModelName{} consistently improves recommendation accuracy across diverse backbones and indexing methods, while substantially reducing both training complexity and inference latency.

\bibliography{main}
\bibliographystyle{iclr2026_conference}

% \begin{table*}[htbp]
%     \centering
%     \caption{Data examples of POI information.}
%     \label{Table_POI_data_examples}
%     \begin{tabular}{cccccc}
%         \toprule
%         \textbf{POI ID} & \textbf{Nscore} & \textbf{GID} & \textbf{CID} & \textbf{ARID} & \textbf{Coordinates} \\
%         \midrule
%         0 & 0.436961 & 25556599 & 8 & 353 & 120.350943,32.177339 \\
%         1 & 0.389202 & 5901606 & 1 & 2089 & 112.355048,33.531655 \\
%         2 & 0.190659 & 17952948 & 17 & 3192 & 115.554458,33.019564 \\
%         \bottomrule
%     \end{tabular}
% \end{table*}

\end{document}